\documentclass[aps,floatfix,twocolumn,amsmath,amssymb,footinbib,superscriptaddress,showpacs]{revtex4-1}
\usepackage{graphics}      
\usepackage{graphicx}      
\usepackage{longtable}     
\usepackage{url}           
\usepackage{bm}
\usepackage{braket}        
\usepackage{xcolor}
\usepackage{float}
\usepackage{natbib}
\usepackage{graphicx}
\usepackage{comment}
\usepackage{siunitx}
\usepackage{appendix}
\usepackage{amsmath,amssymb,epsfig,color}
\usepackage{ulem}
\usepackage{xcolor}
\usepackage{framed}

\newcommand{\Eq}[1]{Eq.\,(\ref{#1})}
\newcommand{\Fig}[1]{Fig.\,\ref{#1}}

\newcommand{\nl}{\nonumber \\}
\newcommand{\be}{\begin{equation}}
\newcommand{\ee}{\end{equation}}
\newcommand{\bea}{\begin{eqnarray}}
\newcommand{\eea}{\end{eqnarray}}

\begin{document}
	
\title {Quantum dissipative dynamics of driven Duffing oscillator near attractors}
\author{Wei Feng}
\affiliation{Center for Joint Quantum Studies and Department of Physics, School of Science, Tianjin University, Tianjin 300072, China}
	
\author{Lingzhen Guo}
\thanks{lingzhen\_guo@tju.edu.cn}
\affiliation{Center for Joint Quantum Studies and Department of Physics, School of Science, Tianjin University, Tianjin 300072, China}
	
\begin{abstract}
We investigate the quantum dissipative dynamics near the stable states (attractors) of a driven
Duffing oscillator. A refined perturbation theory that can treat two perturbative parameters with
different orders is developed to calculate the quantum properties of the Duffing oscillator near the
attractors. We obtain the perturbative analytical results, that go beyond the standard linearization approach, for the renormalized level spacings, the orbital displacements, and the effective temperature near the classical attractor. Furthermore, we demonstrate that strong damping induces additional
slight renormalization of level spacings, and the Bose distribution together with dephasing.
Our work provides new insights into the quantum dynamics of the driven Duffing oscillator and offers
a theoretical framework that can be applied to related quantum systems near their stable states.
\end{abstract}

\date{\today}
\maketitle
	
\section{Introduction}\label{sec-intro}
	
The driven Duffing oscillator \cite{Duf18}, a paradigmatic model for various nonlinear mechanical systems and nonlinear optical phenomena, has fascinated physicists for a long time with its rich dynamical behaviors such as bistability, bifurcation, and chaotic trajectories.
In recent years, the Duffing oscillator has received renewed attention
as the quantum regime of nanomechanical oscillators becomes experimentally accessible.
The interplay between quantum effects, nonequilibrium dynamics,
and nonlinear effects makes the driven Duffing oscillator a model with broad applicability across multiple domains, e.g., mechanical metrology \cite{Dyk12m,Cle02,Lif08,Poo12}, chaotic dynamics \cite{Kro23,Gho23}, cavity and circuit quantum electrodynamics \cite{Zhang13n,Jea24,Mav16,Boi10},
nano- and opto-mechanics \cite{Buk06,Alm07,Del07,Kip11,Dyk22,Ken24,Pet14,Bol23,Hel24,Dyk22R,Mil08}, and cold atoms \cite{Art03,Got19, Par12}.
Notable examples include bifurcation-based quantum measurement devices, where the low- and high-amplitude states of bistability are entangled with the ground and excited states of qubits, respectively,
enabling the analysis of the qubit states through the detection of classical signals \cite{Ald05,Sid04,Sid06,Man07,Sid09}.

In practice, physical systems inevitably interact with their environment,
leading to the decoherence of quantum states and dissipation of energy.
The dissipative dynamics of driven Duffing oscillators have been extensively studied \cite{Chen23New,Mar21New}.
In the underdamped regime near a bifurcation point, a scaling law for the noise-induced escape from metastable states was established \cite{Dyk05}.
Close to the bottom of the well in a parametrically driven Duffing oscillator,  it was identified that the energy dependence of the level spacings captured by the perturbative approach beyond linearization gives rise to a fine structure in the power spectrum \cite{Dyk11n}.
It was also revealed that the quantum activation process has distinct temperature dependency compared to that for the quantum tunneling process \cite{Dyk07}.
The distinct transition rate scaling behaviors near bifurcation points were also revealed in the driven mesoscopic Duffing oscillator \cite{Guo10}.
It was found that the bifurcation point is shifted by the quantum effect and
a linear scaling behavior for the tunneling rate with the driving distance to the shifted bifurcation point \cite{Guo11}.
Recent advances also showed that the quadrature squeezing can enhance the Wigner negativity in a Duffing oscillator, demonstrating a promising approach to generate nonclassical states in macroscopic mechanical systems \cite{Ros24}.
For two Duffing oscillators coupled via nonlinear interactions, the stationary paired solutions and their dynamical stability were demonstrated \cite{Hel24}.
In a coupled system consisting of a time-delayed Duffing oscillator (as a driver system) and a non-delayed Duffing oscillator (as a response system), the phenomenon of transmitted resonance was investigated \cite{Coc24}.

In this paper, we focus on investigating quantum dissipative dynamics near the attractors of a driven Duffing oscillator.
We develop an effective quantum master equation
that can address quantum fluctuations, thermal effects, damping, and dephasing in a unified framework.
Our approach is essential to quantify the occupation of high levels near the bottom of the potential well by the quantum squeezing effects.
We also demonstrate the effects of strong damping and dephasing on the system's dynamics,
including level spacing renormalization and dephasing-modified Bose distributions.
While prior studies within linearized frameworks have successfully captured phenomena such as effective temperatures \cite{Dyk11n,Pea10n,Lem15n},
the refined perturbation theory presented in our work goes beyond the standard linearization approach allowing us to investigate the energy dependence of the level spacing, the orbital displacement, and the effective temperature.
To address this, we develop a refined perturbation theory capable of treating two perturbative parameters with distinct orders,
enabling a unified analysis of nonlinear and dissipative quantum effects near the attractors.
By comparing our theoretical predictions with exact numerical simulations,
we demonstrate the accuracy and utility of our proposed framework.

\clearpage

\section{General Theory}\label{sec-ME}

\subsection{Model Hamiltonian}

An extensive class of macroscopic physical systems,
such as Josephson junctions and nanomechanical oscillators can be modeled by the Duffing oscillator in the presence of a periodic driving force, with the system Hamiltonian described by \cite{Ser07}
\begin{equation}\label{DDO}
{H}_S(t)=\frac{{p}^2}{2m}+\frac{1}{2}m\Omega^2{x}^2
-\gamma{x}^4+F(t){x}.
\end{equation}
Here, parameter $m$ ($\Omega$) describes the mass (frequency) of the oscillator, $\gamma$ gives the nonlinearity of Duffing oscillator, and $F(t)=F_{0}(e^{i\nu t}+e^{-i\nu t})$ describes the periodical driving force with frequency $\nu$.
By switching to the rotating frame using the transformation $U(t) = \exp(-i\nu a^\dagger a t)$ with
$a^\dagger$ ($a$) the raising (lowering) operator, and applying the rotating wave approximation (RWA), we obtain a time-independent Hamiltonian

\begin{equation}\label{H0}
H=(\delta\omega+ \chi) a^\dagger a + \chi (a^\dagger a)^2 + \epsilon(a^\dagger + a). \
\end{equation}
Here, parameter $\delta \omega=\hbar(\Omega-\nu)$ is the frequency detuning,
$\chi=-3\gamma\hbar^2/2(m\Omega)^2$ is the scaled dimensionless nonlinearity, and
$\epsilon=F_0\sqrt{{\hbar}/{2m\Omega}}$ is the scaled driving strength.
We then introduce the position operator $Q$ and momentum operator $P$ in the rotating frame via
\begin{equation}\label{QP}
Q=\sqrt{\frac{\lambda}{2}}(a^\dagger + a),\ \ \
P=i\sqrt{\frac{\lambda}{2}}(a^\dagger - a),
\end{equation}
which satisfy the commutation relation
\begin{equation}
[Q, P]=i\lambda.
\end{equation}
Here, the parameter $\lambda=-\chi/(4\Delta)$ is the dimensionless Planck constant that describes the quantumness of the system, i.e., the value of $\lambda$ increases as the system approaches the quantum regime.
Substituting operators $Q$ and $P$ back into the RWA Hamiltonian (\ref{H0}), we
obtain
\begin{equation}\label{Hc}
H/\Delta=\frac{1}{\lambda}(g+\frac{1}{4})-\frac{1}{2}+\frac{\lambda}{4},
\end{equation}
where $g$ is the quasienergy given by
\begin{equation}\label{gDuffing}
g=-(Q^2+P^2-1)^2/4+\sqrt{\beta}Q.
\end{equation}
Here, the parameter $\beta=-f^2\chi/(2\Delta^3)$ is the scaled driving strength.
Note that Eq.~(\ref{gDuffing}) is valid only for the soft nonlinearity $\chi>0$.
For the hard nonlinearity $\chi<0$, the quasienergy is given by $g=(Q^2+P^2-1)^2/4-\sqrt{\beta}Q$.

\subsection{Renormalized master equation}

\begin{figure}
\centering
\includegraphics[scale=0.75]{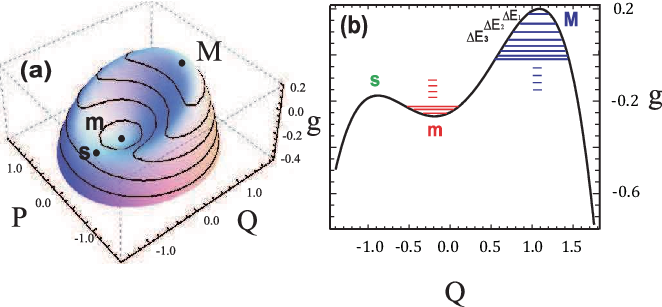}
\caption{
(a) Quasienergy landscape in phase space of a driven Duffing oscillator.
The extrema correspond to the high-amplitude stable state (M) and the low-amplitude stable state (m), while the saddle point (s) marks an unstable state.
%
%
(b) Cross-section of the quasienergy potential at $P=0$.
Quantum energy levels close to the maximum (M) and minimum (m) of the potential are depicted by blue and red lines, respectively.
The unstable saddle point (s) is also indicated.
}\label{Fig-quasienergy}
\end{figure}
The characteristic behavior of the driven Duffing system is the \textit{bistability} manifesting as two stable states: the low-amplitude state (LAS) and the high-amplitude state (HAS).
As depicted in Fig. 1(a), these stable states correspond to the extrema in the quasienergy landscape, which are defined as the \textit{attractors} in the phase space.
The unstable state, known as the saddle point, is located on the separatrix, serving as the boundary dividing the basins of the attractors. When the damping is present, the system evolves towards the nearby attractor if its initial state lies within the basin of the attractor.
However, due to thermal noise, the system does not remain exactly on the attractor, but forms a probability distribution within its basin of attraction.
To describe the dissipative dynamics of our system, we employ the Lindblad form master equation
\begin{equation}\label{ME}
\frac{d\rho}{dt}=-i[H,{\rho}]
+ \frac{\kappa}{2} \{ (1+\bar{n}){\cal D}[a]{\rho}
+ \bar{n}{\cal D}[a^{\dagger}]{\rho} \}   ,
\end{equation}
where ${\cal D}[\bullet]$ is the Lindblad operator defined as
${\cal D}[A]{\rho}\equiv 2A{\rho}A^{\dagger}-A^{\dagger}A{\rho}-{\rho}
A^{\dagger}A $, $\bar{n}$ is the Bose-Einstein distribution, and
$\kappa$ is the damping strength.

To study the quantum dynamics near the bottom of each stable state,vwe first transform the system to the center of the attractor using the displacement operator
\begin{eqnarray}\label{displace}
D[\alpha]=e^{\alpha a^\dagger - \alpha^*a}
\end{eqnarray}
with parameter $\alpha$ a complex number. By defining the displaced density matrix $\tilde{\rho} = D[\alpha]^\dagger \rho D[\alpha]$, we obtain the following master equation
\begin{eqnarray}
\frac{d\tilde{\rho}}{dt} &=& -i \left[ \tilde{H},{\tilde{\rho}}  \right]
+ \frac{\kappa}{2} \{ (1+\bar{n}){\cal D}[a]{\tilde{\rho}}
+ \bar{n}{\cal D}[a^{\dagger}]{\tilde{\rho}} \} \nl
&& + \left[ {\tilde{\alpha}}^*a-{\tilde{\alpha}}a^\dagger,\tilde{\rho} \right]   \,,
\end{eqnarray}
with
$
{\tilde{\alpha}}=[\frac{1}{2}\kappa+i(\delta\omega+\chi+2\chi |\alpha|^2)]\alpha+i\varepsilon.
$ By choosing  $\alpha$ such that $\tilde{\alpha}=0$, the master equation is simplified into
\begin{eqnarray}\label{ME1}
\frac{d\tilde{\rho}}{dt}&=&-[\tilde{H},{\tilde{\rho}}]
+ \frac{\kappa}{2} \{ (1+\bar{n}){\cal D}[a]{\tilde{\rho}}
+ \bar{n}{\cal D}[a^{\dagger}]{\tilde{\rho}} \}
\end{eqnarray}
with the renormalized Hamiltonian $\tilde{H}$ given by
\begin{eqnarray}\label{H1}
\tilde{H}&=&(\delta\omega+4\chi|\alpha|^2) a^\dagger a
+ \chi (a^\dagger a)^2 +\chi ({\alpha^*}^2a^2+{\alpha}^2{a^{\dagger}}^2) \nl
&&+2\chi(\alpha {a^\dagger}^2a + \alpha^*{a^\dagger}a^2).
\end{eqnarray}
We then introduce the squeezing operator
\begin{eqnarray}\label{squeezing}
S(\xi)=e^{(\xi^* a^2 - \xi {a^\dagger}^2)/2},
\end{eqnarray}
which has the transformation property $S^\dagger a S = va + ua^\dagger$ with
$v=\cosh(|\xi|)$ and $u=-\frac{|\xi|}{\xi} \sinh(|\xi|)$.
By defining the squeezed density operator
$\bar{\rho}=S^\dagger \tilde{\rho} S$,
we transform the master equation \Eq{ME1} into the following form
\begin{eqnarray}\label{eq-MEsqu}
\frac{d\bar{\rho}}{dt}&=&-i[\bar{H},\bar{\rho}]
+ \frac{\kappa}{2} \{ (1+\bar{N}){\cal D}[a]\bar{\rho}
+ \bar{N}{\cal D}[a^{\dagger}]\bar{\rho} \} \nl
&&+\frac{\kappa}{2}M(2a^\dagger\bar{\rho}a^\dagger-{a^\dagger}^2\bar{\rho}-\bar{\rho}{a^\dagger}^2)\nl
&&+\frac{\kappa}{2}M^*(2a\bar{\rho}a-{a}^2\bar{\rho}-\bar{\rho}{a}^2)\nl
&&-i[\bar{\xi}{a^\dagger}^2+\bar{\xi}^*{a}^2,\bar{\rho}],
\end{eqnarray}
where
$\bar{N}=\bar{n}|v|^2+(1+\bar{n})|u|^2$ is the effective Bose distribution and
$M=uv^*(2\bar{n}+1)$ is the squeezing number.
The parameter $\bar{\xi}$ in the last term of Eq.~(\ref{eq-MEsqu}) is given by
$$\bar{\xi}=[\delta\omega+4\chi|\alpha|^2+2\chi(2|u|^2+|v|^2)]v^*u+\chi({\alpha^*}^2u^2+\alpha^2{v^*}^2).$$
By setting $\bar{\xi}=0$, the renormalized master equation (\ref{eq-MEsqu}) is further simplified into
\begin{eqnarray}\label{ME22}
\frac{d\bar{\rho}}{dt}&=&-i[\bar{H},\bar{\rho}]
+ \frac{\kappa}{2} \{ (1+\bar{N}){\cal D}[a]\bar{\rho}
+ \bar{N}{\cal D}[a^{\dagger}]\bar{\rho} \}\nl
&&+\frac{\kappa}{2}M(2a^\dagger\bar{\rho}a^\dagger-{a^\dagger}^2\bar{\rho}-\bar{\rho}{a^\dagger}^2)\nl
&&+\frac{\kappa}{2}M^*(2a\bar{\rho}a-{a}^2\bar{\rho}-\bar{\rho}{a}^2)\,.
\end{eqnarray}
The final renormalized Hamiltonian $\bar{H}$ becomes
\begin{eqnarray}\label{H2}
\bar{H}&=&\delta\bar{\omega} a^\dagger a
+\bar{\chi} (a^\dagger
a)^2 \\\nonumber
&&+2\chi S^\dagger (\alpha {a^\dagger}^2a + \alpha^*
{a^\dagger}a^2) S
+\chi F,
\end{eqnarray}
with $F=2(|v|^2+|u|^2)(v^*u{a^\dagger}^3a+vu^*a^\dagger
a^3)+(v^*u)^2{a^\dagger}^4+(u^*v)^2{a}^4$.
The renormalized detuning $\delta\bar{\omega}$ and nonlinearity $\bar{\chi}$ are given by
\begin{eqnarray}\label{}
 \left \{ \begin{array}{lll}
\delta\bar{\omega}/\delta\omega&=&(1-4\lambda|\alpha|^2)(|v|^2+|u|^2)\\
&& -2\lambda(2|uv|^2+|u|^4+{\alpha^*}^2uv+\alpha^2u^*v^*),\
\\
 \bar{\chi}/\delta\omega&=&-\lambda(|u|^4+|v|^4+4|uv|^2)
 \end{array} \right.
\end{eqnarray}
with the displacement parameter $\alpha$,
the squeezing parameters $u$ and $v$ satisfying the following steady equations
\begin{eqnarray}\label{uvl}
 \left \{ \begin{array}{lll}
&0=&[\frac{\kappa}{2\delta\omega}+i(1-\lambda-2\lambda|\alpha|^2)]\alpha-i\sqrt{\frac{\beta}{2\lambda}}, \ \\
&0=&[1-4\lambda|\alpha|^2-\lambda(4|u|^2+2|v|^2)]v^*u \\
&&-\lambda({\alpha^*}^2u^2+\alpha^2{v^*}^2).
 \end{array} \right.
\end{eqnarray}
Here, we have introduced
the dimensionless driving strength
$\beta=2\lambda(\varepsilon/\delta\omega)^2$.

\subsection{Orders of perturbative parameters}

To perform perturbation calculations for the renormalized Hamiltonian (\ref{H2}), we can choose the dimensionless Planck constant $\lambda=-\chi/(4\Delta)$  as the natural choice for the perturbation parameter.
However, since the displacement parameter $\alpha$ is also a function of $\lambda$, it is subtle to properly organize the perturbative terms according to their respective orders. In fact, the Hamiltonian (\ref{H2}) should be written in different forms for different attractors.
The stable state of the driven Duffing oscillator can be approximated as a coherent state $|\alpha\rangle$.
By applying the variational
principle in quantum mechanics, i.e.,
$\partial_\alpha\langle\alpha|H|\alpha\rangle=0$, we obtain two solutions for the steady coherent number:
a smaller one $|\alpha_l|^2\approx \beta/(2\lambda)$ for the LAS and a
larger one $|\alpha_h|^2\approx 1/(2\lambda)$ for the HAS.

For the LAS, we sort the terms in the renormalized Hamiltonian (\ref{H2}) as follows
\begin{eqnarray}\label{Hl}
\frac{\bar{H}^l}{\delta\omega}=h^l_{0}+\lambda h^l_{\lambda}+\beta h^l_{\beta}+\sqrt{\lambda\beta} h^l_{\lambda\beta},
\end{eqnarray}
where the four terms on the right-hand side are given by
\begin{eqnarray}
h^l_{0}&=&(|v|^2+|u|^2)a^\dagger a, \nl
h^l_{\lambda}&=&-2(2|uv|^2+|u|^4)a^\dagger a \nl
&&-(|u|^4+|v|^4+4|uv|^2)(a^\dagger a)^2-F, \nl
h^l_{\beta}&=&-\frac{1}{\beta}[{4\lambda|\alpha_l|^2}(|v|^2+|u|^2)\nl
&&+\lambda(2{\alpha^*_l}^2uv+2\alpha^2_lu^*v^*)]a^\dagger a, \nl
h^l_{\lambda\beta}&=&-2\sqrt{{\lambda}/{\beta}}\  S^\dagger
(\alpha_l {a^\dagger}^2a + \alpha^*_l {a^\dagger}a^2) S \,.
\end{eqnarray}
Together with \Eq{uvl}, we can now perform perturbative
calculations near the bottom of the LAS.
Given that the dimensionless driving strength $\beta$ is also small, we consider the sum of the last three terms in the Hamiltonian \Eq{Hl} as the perturbation term and calculate the desired quantities perturbatively.

For the HAS, as the coherent number $|\alpha_h|^2\approx 1/(2\lambda)$
can be significantly large for $\lambda \ll 1$, a more careful sorting of the terms in the renormalized Hamiltonian \Eq{H2} is needed, along with the steady-state condition \Eq{uvl},
to ensure terms of the same order are kept together.
We introduce $\gamma=\sqrt{\lambda}$ as the perturbation parameter and rewrite the Hamiltonian as
\begin{eqnarray}\label{Hh}
\frac{\bar{H}^h}{\delta\omega}&=&h^h_{0}+\gamma h^h_{1}+\gamma h^h_{2},
\end{eqnarray}
where the three terms on the right-hand side are 
\begin{eqnarray}
h^h_{0}&=&[(1-4\lambda|\alpha_h|^2)(|v|^2+|u|^2) \nl
&&-2\lambda({\alpha^*_h}^2uv+\alpha_h^2u^*v^*)]a^\dagger a  \,, \nl
h^h_{1}&=&-\sqrt{{\lambda}}\  S^\dagger
(2\alpha_h {a^\dagger}^2a +2 \alpha^*_h {a^\dagger}a^2+\alpha_h {a^\dagger} + \alpha^*_h a) S  \,, \nl
h^h_{2}&=&-2(2|uv|^2+|u|^4)a^\dagger a-(|u|^4+|v|^4+4|uv|^2)(a^\dagger a)^2 \nl
&& -F-2(2|u|^2+|v|^2)(v^*u{a^\dagger}^2+u^*v{a}^2)  \,.
\end{eqnarray}
To handle the perturbation orders coherently, we have rearranged the perturbation terms by removing those of order
$\lambda$ from the steady condition \Eq{uvl} and incorporating them into the renormalized Hamiltonian.
The coherent number
$\alpha_h$, $u$ and $v$ for the HAS
are now determined by the revised steady condition
\begin{eqnarray}\label{uvh}
\left \{ \begin{array}{lll}
&&[\frac{\kappa}{2\delta\omega}+i(1-2\lambda|\alpha_h|^2)]\alpha_h-i\sqrt{\frac{\beta}{2\lambda}}=0, \nl
&&(1-4\lambda|\alpha_h|^2)v^*u-\lambda({\alpha^*_h}^2u^2+\alpha_h^2{v^*}^2)=0.
\end{array} \right.
\end{eqnarray}

The behavior near the bottom of the LAS is relatively simple
and can be modeled using a harmonic oscillator.
However, for the HAS, the nonlinear term $\chi (a^\dagger a)^2 \approx \chi |\alpha_h|^2$ becomes prominent, and the oscillator behaves as a highly squeezed coherent state.
In the following sections, we will apply our perturbative method to calculate the crucial quantities related to the HAS of the driven Duffing oscillator,  namely, the level spacings, the orbital displacement, and the effective temperature in the vicinity of the HAS attractor.

\section{Results}

\subsection{Quantum dynamics of HAS}\label{sec-HAS}
The quantum dynamics of the driven Duffing oscillator near the HAS attractor
exhibit a rich interplay among nonlinearity, quantum fluctuations, and thermal noise.
In this section, we discuss the quantum properties of the HAS using the renormalized master equation combined with a refined perturbation theory.

\subsubsection{Level Spacing}
\begin{figure}
\center
\includegraphics[scale=0.85]{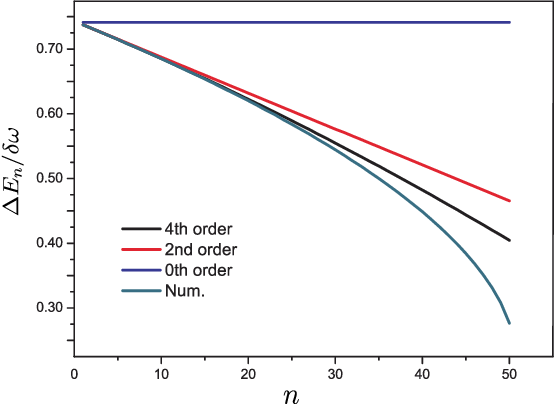}
\caption{
Comparison of perturbation calculations and exact results for energy level spacing.
The exact results (lowest curve) for level spacing $\Delta E_n=|E_{n+1}-E_n|$ are compared with
the zeroth-order (constant spacing), the second-order (second highest line, $\gamma^2=\lambda$) 
and the fourth-order (third highest line, $\gamma^4=\lambda^2$)
corrections. 
Excellent agreement is observed for low-level numbers near the potential well bottom,
while higher-order corrections are required for large $n$.
Parameters: $\lambda=0.016$ and $\beta=4/75$.
}\label{Fig-level1}
\end{figure}

The nonlinear term $\chi (a^\dagger a)^2$ in the Hamiltonian,
having the opposite sign to $\delta \omega$,
results in a decrease in the level spacing $\Delta E_n=|E_{n+1}-E_n|$ as we approach the saddle point,
as illustrated in \Fig{Fig-quasienergy}(b).
One can calculate the level spacings with standard perturbation theory by treating the sum of $\gamma h_1^h$ and $\gamma h_2^h$ in Eq.~(\ref{Hh}) as one perturbation term. However, it becomes a challenge to control the accuracy of the level spacings using the perturbative parameter.
We find it necessary to distinguish between these two perturbative terms in the perturbation calculations to accurately determine the level spacing.
To address this, we have developed a double perturbation theory framework that is particularly suited for the HAS Hamiltonian containing second-order small terms, see the details in Appendix~\ref{app-dpt}.

In \Fig{Fig-level1}, we compare our perturbation calculations with the exact numerical results obtained by diagonalizing the original Hamiltonian \Eq{H0}. These results show an excellent agreement for energy levels near the bottom of the potential well.
Under the zeroth-order perturbation approximation, the energy level spacing remains constant across all levels,
similar to that of the harmonic oscillator.
The second-order and fourth-order corrections provide accuracies up to $\gamma^2=\lambda$ and $\gamma^4=\lambda^2$, respectively.
Higher-order perturbation calculations become necessary for levels farther from the bottom.


\begin{figure}
\center
\includegraphics[scale=0.8]{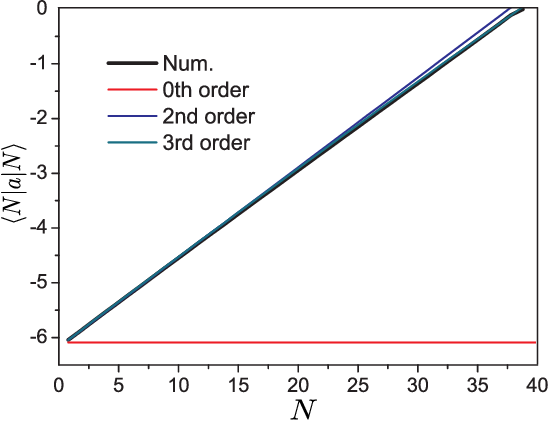}
\caption{
Orbitial displacement.
The average position $\langle N|a|N\rangle$ of the energy level $N$,
illustrates the shift due to the quantum fluctuation.
Under the harmonic approximation, $\langle N|a|N\rangle$ is a constant for every level (lowest line).
Considering higher-order corrections, we observe changes in
$\langle N|a|N\rangle$ across levels.
Compared with the second-order perturbation result (highest line),
the third-order result (second highest line) already aligns well with the exact one (bold line).
}\label{nan}
\end{figure}

\subsubsection{Orbital displacement}
We denote the eigenstate of the renormalized Hamiltonian $\bar{H}$ as $|n'\rangle$, which is generally a superposition of harmonic oscillator eigenstates
$|n'\rangle=|n\rangle+\sum_{k\neq n}\xi_{kn}|k\rangle$, where the superposition coefficients $\xi_{kn}$ are provided in the Appendix~\ref{app-dpt}.
The eigenstate $|N\rangle$ of the original Hamiltonian (\ref{H0}) is related to that of the renormalized Hamiltonian (\ref{H2}) via the relationship $|N\rangle=DS|n'\rangle$.
The matrix element $\langle N|a|M\rangle$ for different levels $|N\rangle$ and $|M\rangle$ is then given by
\begin{eqnarray}\label{eq-NaM}
\langle N|a|M\rangle&=&\langle n'|S^\dagger D^\dagger a D S|m'\rangle \\
&=&\langle n'|(va+u^\dagger+\alpha)|m'\rangle \nl
&=&v\langle n'|a|m'\rangle+u\langle m'|a|n'\rangle^*+\alpha\langle n'|m'\rangle \,. \nonumber
\end{eqnarray}
The matrix element $\langle N|a|N\rangle$ provides insight into the orbital displacement in the phase space.
Under the harmonic approximation ($\xi_{kn}=0$), $\langle N|a|N\rangle$ remains a constant $\alpha$ for all levels.
However, considering higher-order corrections, the orbital displacement $\langle N|a|N\rangle$ changes with energy level, as depicted in Fig.~\ref{nan}.
The perturbation results agree well with numerical calculations.

\begin{figure}
\center
\includegraphics[scale=0.8]{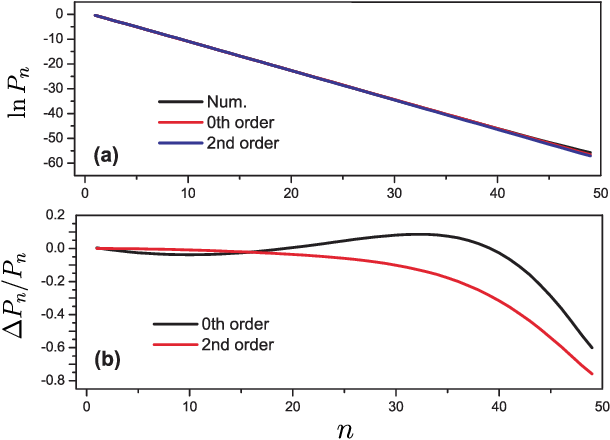}
\caption{Comparison of stationary probability distributions.
(a)
Comparison of the stationary probability distribution obtained
through perturbation theory to zeroth-order (red line), second-order (blue lines)
and exact numerical simulations (black line).
(b) Relative error $\Delta p_n/p_n$ in the stationary distribution
obtained from perturbation theory compared to numerical results.
The discrepancy increases for higher levels but remains less than 1.
}\label{pnN}
\end{figure}

\begin{figure}
\center
\includegraphics[scale=0.8]{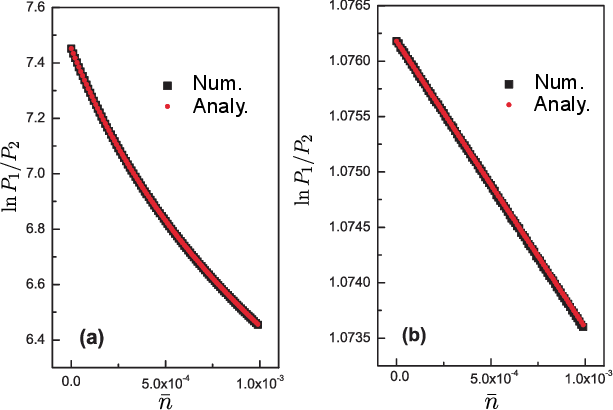}
\caption{
Ratio of probabilities for the lowest levels
in (a) the LAS and (b) HAS, plotted as a function of the Bose distribution $\bar{n}$.
The agreement between the analytical results (black square dots) and the exact numerical results (red circle dots) is excellent.
}\label{Nn}
\end{figure}

\begin{figure}
\center
\includegraphics[scale=0.85]{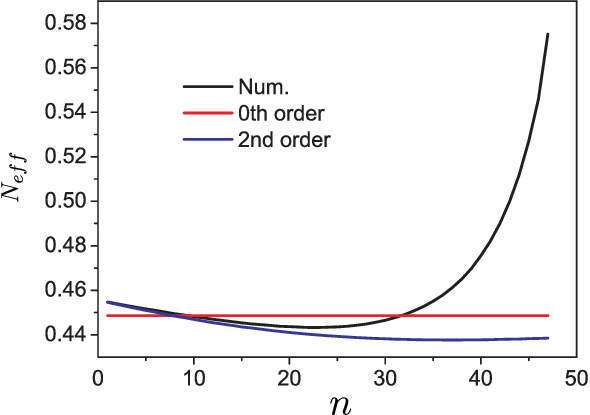}
\caption{Effective temperature of energy levels.
The stationary probability distributions $N_{eff}(n)$ of level $|n\rangle$ obtained
through perturbation theory to zeroth order (red line) and second order (blue lines)
are compared with the exact numerical simulations (black line).
The zero-order effective temperature term gives a constant effective temperature.
The second-order term results in changes in $N_{eff}(n)$,
which are quite accurate for levels near the bottom.
}\label{Neff}
\end{figure}

\subsubsection{Effective temperature}
Next, we calculate the stationary distribution over the levels of the HAS
and the effective temperature near the bottom.
It is important to note that the annihilation operator $a$ is for the Fock
state $|N\rangle$, which decreases the Fock state from a higher level to the next lower level $a|N\rangle= \sqrt{N}|N-1\rangle$.
However, in our case,
the eigenstate of quasienergy $|n\rangle$ is the superposition of 
Fock states $|N\rangle$. As a result, the annihilation operator
$a$ can either decrease or increase the state $|n\rangle$ even
at zero temperature.

Under the assumption of weak damping ($\kappa \ll E_n-E_{n+1}$), the
off-diagonal matrix elements on the state $|n\rangle$ are very small. Thus, we can only keep the diagonal elements.
Here, we assume that the stationary density matrix is diagonal and denote the diagonal terms as 
$p_{n'}=\langle n'|\rho|n'\rangle$.
The master equation (\ref{H2}) can be simplified into a
balance equation \cite{Dyk06}
\begin{equation}\label{pn}
\frac{dp_{n'}}{dt}=\kappa \sum_{m'}(W_{n',m'}p_{m'}-W_{m',n'}p_{n'}),
\end{equation}
where the transition rate from level $|m'\rangle$ to level $|n'\rangle$ ($m'\neq n'$) is given by
\begin{eqnarray}\label{Wnm}
W_{n',m'}&=&M\langle m'|a|n'\rangle^*\langle n'|a|m'\rangle^*
+M^*\langle m'|a|n'\rangle\langle n'|a|m'\rangle \nl
&&+(1+\bar{N})|\langle n'|a|m'\rangle|^2+\bar{N}|\langle m'|a|n'\rangle|^2.
\end{eqnarray}
One can prove that
the transition rate $W_{n',m'}$ for $n'\neq m'$ is equal to $|\langle N|a|M\rangle|^2$ calculated above in \Eq{eq-NaM} .

As can be seen from the above rate equation, even at zero temperature,
the oscillator can make transitions to both lower and higher energy levels.
In Fig.~\ref{pnN}, we compare the stationary distribution obtained using our double perturbation theory with exact numerical results.
To the lowest order, all superposition coefficients are zero.
To the first order of $\sqrt{\lambda}$ (i.e., $\xi_{kn}=\sqrt{\lambda}\xi^{(1)}_{kn}$),
there is no correction to the stationary distribution $p_{n'}$.
Then, to the second order (i.e., $\xi_{kn}=\sqrt{\lambda}\xi^{(1)}_{kn}+\lambda\xi^{(2)}_{kn}$),
we solve the balance equation (\ref{pn}) accordingly.
Fig.~\ref{pnN} (b) illustrates the relative error $\Delta p_n/p_n$ in comparison to the exact numerical results, which shows that the discrepancy for low levels is mitigated by high-order perturbative calculations.

In the vicinity of the bottom, we can apply the harmonic approximation ($ \xi_{kn} = 0$).
Under this approximation, the ratio of probabilities over adjacent levels is
\begin{eqnarray} \label{Bose}
\frac{p_{n'+1}}{p_{n'}}&=&\frac{W_{n'+1,n'}}{W_{n',n'+1}} =\frac{\bar{N}}{1+\bar{N}}\nl
&=&\frac{\bar{n}|v|^2+(1+\bar{n})|u|^2}{1+\bar{n}|v|^2+(1+\bar{n})|u|^2}.
\end{eqnarray}
We verify the relationship between $\ln(p_1/p_2)$ and the Bose distribution $\bar{n}$ in \Fig{Nn}.
The agreement between the analytical results from \Eq{Bose} and the exact numerical one is excellent.

For high levels, we can define the level-dependent effective temperature via $$N_{eff}(n) \equiv p_{n+1}/(p_{n}-p_{n+1})$$ for level $|n\rangle$.
Fig.~\ref{Neff} illustrates how $N_{eff}$ varies from the lowest to higher levels.
The zero-order term yields a constant effective temperature.
When we include the correction up to the order of $\lambda$, the correction leads to changes in $N_{eff}$ showing a good agreement with the numerical results for levels near the bottom.

\begin{figure}
\center
\includegraphics[scale=0.8]{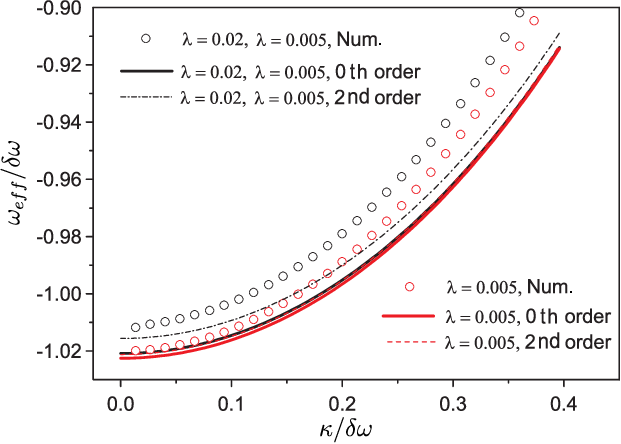}
\caption{
Relationship between effective frequency $\omega_{eff}=\Delta E_1$ and damping strength $\kappa$.
Results obtained from the emission spectrum method (circles), cf. Eq.~(\ref{eq-sw}), are compared with those from our perturbation theory (solid and dashed lines).
}\label{damping}
\end{figure}

\begin{figure}
\center
\includegraphics[scale=0.7]{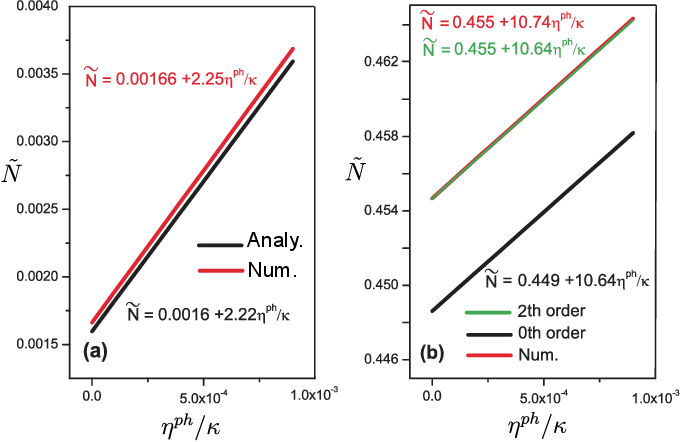}
\caption{
Renormalized effective Bose distribution $\tilde{N}$ due to dephasing effects for (a) the LAS and (b) the HAS.
We compare theoretical predictions from \Eq{eq-reBose} with the exact numerical results and extract the linear relationships.
}\label{dephasing}
\end{figure}

\subsection{Strong Damping and Dephasing}\label{sec-damping}
In this section, we explore the dynamics and the stationary state of the system
under conditions of strong damping and dephasing,
which can significantly alter the behavior predicted by the weak damping approximation.

\subsubsection{Strong damping}
In the regime of strong damping, the harmonic approximation,
which leads to the level spacing $\Delta E_n$ that is independent of damping, is
no longer valid. Instead, the level spacing undergoes slight renormalization for strong damping. 
Such an effect can be observed through the emission spectrum  $S(\omega)$,
which represents the spectral density of photons emitted by the driven resonator and is given by
\begin{eqnarray}\label{eq-sw}
S(\omega)
= 2 \operatorname{Re} \int_{0}^{\infty}  e^{-i \omega t}
\text{Tr}[a^{\dagger} e^{\mathcal{L}t} (a \rho_{\text{st}})]  {\rm d}t \,.
\end{eqnarray}
Our method inherently incorporates damping effects into the calculation of $\Delta E_n$.
For finite damping, the squeezing parameters $u$ and $v$ are complex numbers, determined by the steady-state conditions given in \Eq{uvh}.
Substituting these parameters into \Eq{Hh},
we obtain the effective frequency $\omega_{eff}=\Delta E_1$.
In \Fig{damping}, we compare the results obtained through the emission spectrum Eq.~(\ref{eq-sw}) with those from our perturbation theory,
demonstrating satisfactory consistency.
The minor discrepancy arises primarily from higher-level spacings $\Delta E_n (n>1)$,
which are generally smaller than the first-level spacing $\Delta E_1$.
For more accurate results, the average of all level spacings should be considered in the calculations.

\subsubsection{Dephasing}
To incorporate dephasing, we introduce the dephasing term $\eta^{ph}{\cal D}[a^{\dagger}a]{\rho}$
into the master equation (\ref{ME}).
For convenience, we define a generalized Lindblad operator ${\cal L}[A;B]\rho\equiv2A\rho B -BA\rho-\rho AB$.
In the spirit of the rotating wave approximation,
we obtain the renormalized master equation for the displaced and squeezed density operator $\bar{\rho}=S^\dagger D^\dagger{\rho}DS$ (see the detailed derivation in Appendix~\ref{app-deph})
\begin{eqnarray}\label{ME2}
\frac{d\bar{\rho}}{dt}&=&-i[\bar{H},\bar{\rho}]
+ \frac{\kappa}{2} \{ (1+\tilde{N}){\cal D}[a]\bar{\rho}
+ \tilde{N}{\cal D}[a^{\dagger}]\bar{\rho} \}\nl
&&
+\frac{\kappa}{2}M{\cal L}[a^\dagger;a^\dagger]{\bar{\rho}}
+\frac{\kappa}{2}M^*{\cal L}[a;a]{\bar{\rho}} \nl
&&+\eta^{ph}(|v|^2+|u|^2)^2{\cal D}[a^{\dagger}a]{\bar{\rho}}\nl
&&+\eta^{ph}|uv|^2({\cal D}[{a^{\dagger}}^2]{\bar{\rho}}+{\cal D}[a^2]{\bar{\rho}}),
\end{eqnarray}
where the renormalized Bose distribution, affected by dephasing, is given by:
\begin{eqnarray}\label{eq-reBose}
\tilde{N}&=&\bar{N}+\frac{\eta^{ph}}{\kappa}|\alpha^*u+\alpha v^*|^2 \nl
&=&|u|^2+\bar{n}(|u|^2+|v|^2)+\frac{\eta^{ph}}{\kappa}|\alpha^*u+\alpha v^*|^2 \,.
\end{eqnarray}
We verify our predictions for the renormalized Bose distribution
by comparing them with exact numerical simulations according to the probabilities over the two lowest levels,
specifically, $\tilde{N}=p_2/(p_1-p_2)$.
In \Fig{dephasing}, we plot and extract the renormalized Bose distribution as a function of dephasing, which demonstrates an excellent agreement
between the numerical results and the analytical predictions given by \Eq{eq-reBose}
for both LAS and HAS.

\section{Conclusions}\label{sec-sum}

In this work, we have investigated the quantum dissipative dynamics of a driven Duffing oscillator near the bottoms of its stable states.
We elucidated the intricate interplay among the nonlinearity, quantum fluctuations, and the influence of an external driving field. We formulated an effective quantum master equation that encompasses quantum and thermal fluctuations, strong damping, and dephasing within a unified framework.
We have developed a refined perturbation approach to analyze the quantum dynamics near both the LAS and the HAS of the Duffing oscillator.
While the LAS behavior can be approximated using a harmonic oscillator model,
the HAS exhibits more complex behavior owing to significant nonlinear terms.
Because of quantum fluctuations, even at zero temperature, higher energy levels near the bottom of the potential well are excited.
We calculated the level spacing and effective temperature near the bottom of the HAS and compared them with numerical simulations, demonstrating the accuracy and utility of our proposed approach.

We also investigated the effects of strong damping and dephasing on the system's dynamics.
We showed that the level spacing undergoes slight renormalization for strong damping,
which can be observed through the emission spectrum.
We derived the renormalized quantum master equation and analyzed the system's behavior affected by dephasing.
Our work provides new insights into the quantum dynamics of driven Duffing oscillators, particularly near their stable states, and offers a theoretical framework that can be applied to related quantum systems under strong damping and dephasing conditions.

\bigskip
{\bf ACKNOWLEDGMENTS}

This work was supported by the Natural Science Foundation of China (Grant No. 12475025).

{\bf AUTHOR DECLARATIONS}

{\bf Conflict of interest}

The authors have no conflicts to disclose.

{\bf Author contributions}
%

The two authors contributed equally to this work.

{\bf DATA AVAILABILITY}

The data that support the findings of this study are available within the article.

\onecolumngrid

\appendix

\section{Double Perturbation Theory}\label{app-dpt}

To effectively address Hamiltonians containing second-order perturbation terms,
exemplified by the HAS in the driven Duffing oscillator, we develop a framework for double perturbation theory.
This theory is specifically designed for systems where treating the second-order terms independently is essential for maintaining computational precision and obtaining physically meaningful results.
Consider a Hamiltonian of the general form:
\begin{equation}\label{H}
{H}=H_0+\gamma H_1 + \gamma^2 H_2,
\end{equation}
where $\gamma$ is a small parameter.
In conventional perturbation theory,
the terms $\gamma H_1 + \gamma^2 H_2$ are often treated as a single perturbation term.
However, for the HAS of the driven Duffing system,
this approach does not yield results with the necessary accuracy.
Therefore, we introduce the concept of double perturbation theory, where the terms are handled separately.

The eigenvalues and eigenstates of $H_0$ are denoted as $\epsilon^{(0)}_n$ and
$|\psi^{(0)}_n\rangle$ respectively, satisfying
$H_0|\psi^{(0)}_n\rangle=\epsilon^{(0)}_n|\psi^{(0)}_n\rangle$.
The exact eigenvalues and eigenstates of $H$ are denoted as $\epsilon_n$ and
$|\psi_n\rangle$, which can be expanded in powers of $\gamma$ as:

\begin{eqnarray}
\epsilon_n&=&\epsilon^{(0)}_n+\gamma\epsilon^{(1)}_n+\gamma^2\epsilon^{(2)}_n
+\gamma^3\epsilon^{(3)}_n+\gamma^4\epsilon^{(4)}_n+o(\gamma^5), \nl
|\psi_n\rangle&=&|\psi^{(0)}_n\rangle
+\gamma\sum_{k\neq n}\xi^{(1)}_{kn}|\psi^{(0)}_k\rangle
+\gamma^2\sum_{l\neq n}\xi^{(2)}_{ln}|\psi^{(0)}_l\rangle \nl
&&+\gamma^3\sum_{m\neq n}\xi^{(3)}_{mn}|\psi^{(0)}_m\rangle
+\gamma^4\sum_{p\neq n}\xi^{(4)}_{pn}|\psi^{(0)}_p\rangle+o(\gamma^5).\nl
\end{eqnarray}
From the eigenvalue equation $H|\psi_n\rangle=\epsilon_n|\psi_n\rangle$, we can derive the perturbative results order by order:

1) to the order of $\gamma^0=1$:
$H_0|\psi^{(0)}_n\rangle=\epsilon^{(0)}_n|\psi^{(0)}_n\rangle$;

2) to the order of $\gamma$:
\begin{eqnarray}
&&\gamma(\sum_{k\neq n}\epsilon^{(0)}_k\xi^{(1)}_{kn}|\psi^{(0)}_k\rangle+H_1|\psi^{(0)}_n\rangle) \nl
&=&\gamma(\sum_{k\neq n}\epsilon^{(0)}_n\xi^{(1)}_{kn}|\psi^{(0)}_k\rangle+\epsilon^{(1)}_n|\psi^{(0)}_n\rangle) \,,
\end{eqnarray}
which gives us the perturbative result to the first order
\begin{eqnarray}
\epsilon^{(1)}_n&=&\langle\psi^{(0)}_n|H_1|\psi^{(0)}_n\rangle,\nonumber\\
\xi^{(1)}_{kn}&=&\frac{1}{\epsilon^{(0)}_n-\epsilon^{(0)}_k}\langle\psi^{(0)}_k|H_1|\psi^{(0)}_n\rangle;
\end{eqnarray}

3) to the order of $\gamma^2$:
\begin{eqnarray}
&&\gamma^2(\sum_{l\neq n}\epsilon^{(0)}_l\xi^{(2)}_{ln}|\psi^{(0)}_l\rangle
+\sum_{k\neq n}\xi^{(1)}_{kn}H_1|\psi^{(0)}_k\rangle+H_2|\psi^{(0)}_n\rangle) \nl
&&=\gamma^2(\sum_{l\neq n}\epsilon^{(0)}_n\xi^{(2)}_{ln}|\psi^{(0)}_l\rangle
+\sum_{k\neq n}\xi^{(1)}_{kn}\epsilon^{(1)}_n|\psi^{(0)}_k\rangle+\epsilon^{(2)}_n|\psi^{(0)}_n\rangle), \nl
\end{eqnarray}
which gives us the second-order perturbative result:
\begin{eqnarray}
\epsilon^{(2)}_n&=&\sum_{k\neq n}\langle\psi^{(0)}_n|H_1|\psi^{(0)}_k\rangle \xi^{(1)}_{kn}+\langle\psi^{(0)}_n|H_2|\psi^{(0)}_n\rangle,\nonumber\\
\xi^{(2)}_{ln}&=&\frac{1}{\epsilon^{(0)}_n-\epsilon^{(0)}_l}(\sum_{k\neq n}\langle\psi^{(0)}_l|H_1|\psi^{(0)}_k\rangle
\xi^{(1)}_{kn}+\langle\psi^{(0)}_l|H_2|\psi^{(0)}_n\rangle \nl
&&-\epsilon^{(1)}_n\xi^{(1)}_{ln});
\end{eqnarray}

4) to the order of $\gamma^3$:
\begin{eqnarray}
&&\gamma^3(\sum_{m\neq
n}\epsilon^{(0)}_m\xi^{(3)}_{mn}|\psi^{(0)}_m\rangle+\sum_{l\neq n}\xi^{(2)}_{ln}H_1|\psi^{(0)}_l\rangle \nl
&&+\sum_{k\neq n}\xi^{(1)}_{kn}H_2|\psi^{(0)}_k\rangle) \nl
&=&\gamma^3(\sum_{m\neq n}\epsilon^{(0)}_n\xi^{(3)}_{mn}|\psi^{(0)}_m\rangle
+\sum_{l\neq n}\epsilon^{(1)}_n\xi^{(2)}_{ln}|\psi^{(0)}_l\rangle \nl
&&+\sum_{k\neq n}\xi^{(1)}_{kn}\epsilon^{(2)}_n|\psi^{(0)}_k\rangle
+\epsilon^{(3)}_n|\psi^{(0)}_n\rangle),
\end{eqnarray}

which gives us the perturbative result to the third order:
\begin{eqnarray}
\epsilon^{(3)}_n&=&\sum_{l\neq n}(\langle\psi^{(0)}_n|H_1|\psi^{(0)}_l\rangle \xi^{(2)}_{ln}
+\langle\psi^{(0)}_n|H_2|\psi^{(0)}_l\rangle\xi^{(1)}_{ln}) \,, \nonumber\\
\xi^{(3)}_{mn}&=&\frac{1}{\epsilon^{(0)}_n-\epsilon^{(0)}_m}
[\sum_{l\neq n}(\langle\psi^{(0)}_m|H_1|\psi^{(0)}_l\rangle \xi^{(2)}_{ln} \nl
&&+\langle\psi^{(0)}_m|H_2|\psi^{(0)}_l\rangle\xi^{(1)}_{ln})-\epsilon^{(1)}_n\xi^{(2)}_{mn}-\epsilon^{(2)}_n\xi^{(1)}_{mn}];
\end{eqnarray}

5) to the order of $\gamma^4$:
\begin{eqnarray}
&&\gamma^4(\sum_{p\neq n}\epsilon^{(0)}_p\xi^{(4)}_{pn}|\psi^{(0)}_p\rangle
+\sum_{m\neq n}\xi^{(3)}_{mn}H_1|\psi^{(0)}_m\rangle
+\sum_{l\neq n}\xi^{(2)}_{ln}H_2|\psi^{(0)}_l\rangle)\nl
&&=\gamma^4[\sum_{m\neq n}(\epsilon^{(0)}_n\xi^{(4)}_{mn}
+\epsilon^{(1)}_n\xi^{(3)}_{mn}
+\xi^{(2)}_{mn}\epsilon^{(2)}_n
+\xi^{(1)}_{mn}\epsilon^{(3)}_n)|\psi^{(0)}_m\rangle \nl
&&+\epsilon^{(4)}_n|\psi^{(0)}_n\rangle],
\end{eqnarray}
which gives us the fourth order perturbative result:
\begin{eqnarray}
\epsilon^{(4)}_n&=&\sum_{l\neq n}(\langle\psi^{(0)}_n|H_1|\psi^{(0)}_l\rangle \xi^{(3)}_{ln}+\langle\psi^{(0)}_n|H_2|\psi^{(0)}_l\rangle\xi^{(2)}_{ln}), \nl
\xi^{(4)}_{pn}&=&\frac{1}{\epsilon^{(0)}_n-\epsilon^{(0)}_p}[\sum_{l\neq n}(\langle\psi^{(0)}_p|H_1|\psi^{(0)}_l\rangle \xi^{(3)}_{ln}
+\langle\psi^{(0)}_p|H_2|\psi^{(0)}_l\rangle\xi^{(2)}_{ln}) \nl
&&-\epsilon^{(1)}_n\xi^{(3)}_{pn}-\epsilon^{(2)}_n\xi^{(2)}_{pn}-\epsilon^{(3)}_n\xi^{(1)}_{pn}].
\end{eqnarray}
Similar expressions can be derived for higher-order corrections. \\

For the high-amplitude state of the driven Duffing oscillator,
the perturbative Hamiltonian is given by \Eq{Hh},
\begin{eqnarray}
\frac{\bar{H}^h}{\delta\omega}&=&h^h_{0}+\gamma h^h_{1}+\gamma h^h_{2}, \nonumber\\
\end{eqnarray}
where the order parameter $\gamma$ is $\sqrt{\lambda}$.
The zeroth-order Hamiltonian
\begin{eqnarray}
h^h_{0}&=&[(1-4\lambda|\alpha_h|^2)(|v|^2+|u|^2) \nl
&&-2\lambda({\alpha^*_h}^2uv+\alpha_h^2u^*v^*)]a^\dagger a ,
\end{eqnarray}
corresponds to a harmonic oscillator, and the eigenstates $|\psi^{(0)}_k\rangle$
are simply the harmonic oscillator states $|k\rangle$ with eigenvalues:
\begin{eqnarray}
\epsilon^{(0)}_k&=&k[(1-4\lambda|\alpha_h|^2)(|v|^2+|u|^2) \nl
&&-2\lambda({\alpha^*_h}^2uv+\alpha_h^2u^*v^*)].
\end{eqnarray}

By calculating the matrix elements $\langle l|h^h_{1}|k\rangle$ and $\langle l|h^h_{2}|k\rangle$,
\onecolumngrid
\begin{eqnarray}
\langle l|h^h_{1}|k\rangle&=&-2\sqrt{\lambda
k}[\alpha_h|v|^2u^*(2k-1)+\alpha_h|u|^2u^*(k+1)+\alpha_h^*v|u|^2(2k+1)+\alpha_h^*v|v|^2(k-1)+(\alpha_h^*v+\alpha_hu^*)/2]\delta_{l,k-1}\nonumber\\
&&-2\sqrt{\lambda
(k+1)}[\alpha_h|u|^2v^*(2k+3)+\alpha_h|v|^2v^*k+\alpha_h^*u|v|^2(2k+1)+\alpha_h^*u|u|^2(k+2)+(\alpha_h^*u+\alpha_hv^*)/2]\delta_{l,k+1}\nonumber\\
&&-2u^*v\sqrt{\lambda
k(k-1)(k-2)}(\alpha_h^*v+\alpha_hu^*)\delta_{l,k-3}-2v^*u\sqrt{\lambda
(k+1)(k+2)(k+3)}(\alpha_h^*u+\alpha_hv^*)\delta_{l,k+3},\nonumber\\
\langle
l|h^h_{2}|k\rangle&=&-[2(2|uv|^2+|u|^4)k+(|u|^4+|v|^4+4|uv|^2)k^2]\delta_{l,k}\nonumber\\&&-2u^*v\sqrt{
k(k-1)}[k|u|^2+(k-1)|v|^2]\delta_{l,k-2}-2v^*u\sqrt{
(k+1)(k+2)}[(k+2)|u|^2+(k+1)|v|^2]\delta_{l,k+2}\nonumber\\
&&-(u^*v)^2\sqrt{ k(k-1)(k-2)(k-3)}\delta_{l,k-4}-(v^*u)^2\sqrt{
(k+1)(k+2)(k+3)(k+4)}\delta_{l,k+4},
\end{eqnarray}
we can apply the double perturbation theory to obtain the desired perturbative results, such as level spacing and effective temperature, for the high-amplitude state as detailed in the main text.

\section{Dephasing}\label{app-deph}
This section presents the derivation of \Eq{ME2} in the main text.
To incorporate dephasing, we introduce the dephasing term $\eta^{ph}{\cal D}[a^{\dagger}a]{\rho}$
into the master equation.
For convenience, we define a generalized Lindblad operator ${\cal L}[A;B]\rho\equiv2A\rho B -BA\rho-\rho AB$,
which satisfies
\begin{eqnarray}
&&{\cal D}[A]{\rho}={\cal L}[A;A^\dagger]\rho,\nl
&&{\cal D}[A+B]{\rho}={\cal D}[A]{\rho}+{\cal D}[B]{\rho}
+{\cal L} [A;B^\dagger]\rho+{\cal L}[B;A^\dagger]\rho, \nl
&&{\cal L}[A;B+C]\rho={\cal L}[A;B]\rho+{\cal L}[A;C]\rho,\nl
&&{\cal L}[A+B;C]\rho={\cal L}[A;C]\rho+{\cal L}[B;C]\rho, \nl
&&({\cal L}[A;B]\rho)^\dagger={\cal L}[B^\dagger;A^\dagger]\rho.
\end{eqnarray}
Utilizing these properties, we derive the transformed dephasing term under the action of the displacement and squeezing operators:
\begin{eqnarray}\label{SDD}
&&S^\dagger D^\dagger({\cal D}[a^{\dagger}a]{\rho})DS \nl
&&=(|v|^2+|u|^2)^2{\cal D}[a^{\dagger}a]{\bar{\rho}} \
+|\alpha^*u+\alpha v^*|^2({\cal D}[a^{\dagger}]{\bar{\rho}} \nl
&&+{\cal D}[a]{\bar{\rho}})+|uv|^2({\cal D}[{a^{\dagger}}^2]{\bar{\rho}}
+{\cal D}[a^2]{\bar{\rho}})+{\cal L}\bar{\rho}+({\cal L}\bar{\rho})^\dagger.
\end{eqnarray}
The term ${\cal L}\bar{\rho}$ includes various higher-order terms:
%
\begin{eqnarray}
{\cal L}\bar{\rho}&=&|uv|^2{\cal L}[{a^{\dagger}}^2;{a^{\dagger}}^2]{\bar{\rho}}
+|\alpha^*u+\alpha v^*|^2{\cal L}[{a^{\dagger}};{a^{\dagger}}]{\bar{\rho}}
+v^*u(\alpha^*u+\alpha v^*){\cal L}[{a^{\dagger}}^2;{a^{\dagger}}]{\bar{\rho}}
+vu^*(\alpha u^*+\alpha^* v){\cal L}[a;a^2]{\bar{\rho}} \nl
&&+v^*u(\alpha u^*+\alpha^* v){\cal L}[a;{a^{\dagger}}^2]{\bar{\rho}}
+vu^*(\alpha^*u+\alpha v^*){\cal L}[a^2;a^\dagger]{\bar{\rho}}
+(|v|^2+|u|^2)\{v^*u{\cal L}[a^{\dagger}a;{a^{\dagger}}^2]{\bar{\rho}}
+vu^*{\cal L}[a^\dagger a;a^2]{\bar{\rho}} \nl
&&+(\alpha^*u+\alpha v^*){\cal L}[a^\dagger;a^\dagger a]{\bar{\rho}}
+(\alpha u^*+\alpha^* v){\cal L}[a;a^\dagger a]{\bar{\rho}}\}. \nl
\end{eqnarray}
In the spirit of the rotating wave approximation,
we neglect the terms ${\cal L}\bar{\rho}$ and $({\cal L}\bar{\rho})^\dagger$ in \Eq{SDD}.
Adding the dephasing term to the renormalized master equation, we obtain:
\begin{eqnarray}
\frac{d\bar{\rho}}{dt}&=&-i[\bar{H},\bar{\rho}]
+ \frac{\kappa}{2} \{ (1+\tilde{N}){\cal D}[a]\bar{\rho}
+ \tilde{N}{\cal D}[a^{\dagger}]\bar{\rho} \}
+\frac{\kappa}{2}M{\cal L}[a^\dagger;a^\dagger]{\bar{\rho}}
+\frac{\kappa}{2}M^*{\cal L}[a;a]{\bar{\rho}} \nl
&&+\eta^{ph}(|v|^2+|u|^2)^2{\cal D}[a^{\dagger}a]{\bar{\rho}}
+\eta^{ph}|uv|^2({\cal D}[{a^{\dagger}}^2]{\bar{\rho}}+{\cal D}[a^2]{\bar{\rho}}),
\end{eqnarray}
where the renormalized Bose distribution, affected by dephasing, is given by:
\begin{eqnarray} 
\tilde{N}&=&\bar{N}+\frac{\eta^{ph}}{\kappa}|\alpha^*u+\alpha v^*|^2 \nl
&=&|u|^2+\bar{n}(|u|^2+|v|^2)+\frac{\eta^{ph}}{\kappa}|\alpha^*u+\alpha v^*|^2 \,.
\end{eqnarray}

%


\begin{thebibliography}{99}
\bibitem{Duf18}
G. Duffing,
Erzwungene Schwingungen bei ver\"{a}nderlicher Eigenfrequenz und ihre technische Bedeutung,
Ingenieur-Archiv \textbf{8}, 445 (1918).

\bibitem{Dyk12m}
M. I. Dykman,
Fluctuating nonlinear oscillators: from nanomechanics to quantum superconducting circuits,
Oxford University Press, 2012.

\bibitem{Cle02}
A. N. Cleland and M. L. Roukes,
Noise processes in nanomechanical resonators,
J. App. Phys., \textbf{92}, 2758 (2002).

\bibitem{Lif08}
R. Lifshitz and M. C. Cross,
Nonlinear dynamics of nanomechanical and micromechanical resonators,
Wiley Meinheim Press, 2008.

\bibitem{Poo12}
M. Poot and H. S. J. van der Zant,
Mechanical systems in the quantum regime,
Phys. Rep., \textbf{511}, 273 (2012).


\bibitem{Kro23}
K. A. Krok, A. P. Durajski, R. Szcz\c{e}\'{s}niak,
The Abraham-Lorentz force and the time evolution of a chaotic system: The case of charged classical and quantum Duffing oscillators, Chaos \textbf{32}, 073130 (2022).

\bibitem{Gho23}
P. K Ghosh and P. Roy,
On regular and chaotic dynamics of a non-PT-symmetric Hamiltonian system of a coupled Duffing oscillator with balanced loss and gain,
J. Phys. A: Math. Theor. \textbf{53}, 475202 (2020).


\bibitem{Zhang13n}
J. Zhang, Y.-X. Liu ,  \c{S}. K.  \"{O}zdemir, R.-B. Wu, F. Gao, X.-B.Wang, L. Yang, and F. Nori,
Quantum internet using code division multiple access,
Sci. Rep. \textbf{3}, 2211 (2013).
	
\bibitem{Jea24}
J. Choi, H. Hwang, and E. Kim,
Measurement-induced bistability in the excited state of a transmon,
Phys. Rev. Appl. \textbf{22}, 054069 (2024).

\bibitem{Mav16}
T. K. Mavrogordatos, G. Tancredi, M. Elliott, M. J. Peterer, A. Patterson, J. Rahamim, P. J. Leek, E. Ginossar, and M. H. Szyma\'{n}ska,
Simultaneous bistability of a qubit and resonator in circuit quantum electrodynamics,
Phys. Rev. Lett. \textbf{118}, 040402 (2017).
\bibitem{Boi10}
M. Boissonneault, J. M. Gambetta, and A. Blais,
Improved superconducting qubit readout by qubit-induced nonlinearities,
Phys. Rev. Lett. \textbf{105}, 100504 (2010).


\bibitem{Del07}
P. Del'Haye, A. Schliesser, O. Arcizet, T. Wilken, R. Holzwarth, and T. J. Kippenberg,
Optical Frequency Comb Generation from a Monolithic Microresonator,
Nature (London) \textbf{450}, 1214 (2007).

\bibitem{Kip11}
T. J. Kippenberg, R. Holzwarth, and S. A. Diddams,
Microresonator-Based Optical Frequency Combs,
Science \textbf{332}, 555 (2011).

\bibitem{Pet14}
P. D. Drummond and M. Hillery,
The quantum theory of nonlinear optics,
Cambridge University Press, 2014.

\bibitem{Bol23}
E. Bolandhemmat and F. Kheirandish,
Quantum dynamics of a driven parametric oscillator in a Kerr medium,
Sci. Rep.,  \textbf{13}, 9056 (2023).

\bibitem{Hel24}
F. Hellbach, D. De Bernardis, M. Saur, I. Carusotto, W. Belzig, and G. Rastelli,
Nonlinearity-induced symmetry breaking in a system of two parametrically driven Kerr-Duffing oscillators,
New J. Phys.,  \textbf{26}, 103020 (2024).


\bibitem{Ken24}
D. R. Kenigoule Massembele, P. Djorw\'e, A. K. Sarma, A.-H. Abdel-Aty, and S. G. Nana Engo,
Quantum entanglement assisted via Duffing nonlinearity,
Phys. Rev. A \textbf{110}, 043502 (2024).


\bibitem{Buk06}
E. Buks and B. Yurke,
Mass detection with a nonlinear nanomechanical resonator,
Phys. Rev. E \textbf{74}, 046619 (2006).

\bibitem{Alm07}
R. Almog, S. Zaitsev, O. Shtempluck, and E. Buks,
Noise squeezing in a nanomechanical Duffing resonator,
Phys. Rev. Lett. \textbf{98}, 078103 (2007).

\bibitem{Dyk22}
J. S. Ochs, D. K. J. Bone$\rm{\beta}$, G. Rastelli, M. Seitner, W. Belzig , M. I. Dykman , and E. M. Weig,
Frequency Comb from a Single Driven Nonlinear Nanomechanical Mode,
Phys. Rev. X \textbf{12}, 041019 (2022).

\bibitem{Dyk22R}
A. Bachtold, J. Moser, and M. I. Dykman,
Mesoscopic physics of nanomechanical systems,
Rev. Mod. Phys. \textbf{94}, 045005 (2022).

\bibitem{Mil08}
E. Babourina-Brooks, A. Doherty and G. J. Milburn,
Quantum noise in a nanomechanical Duffing resonator,
New J. Phys. \textbf{10} 105020 (2008).


\bibitem{Art03}
R. Artuso and L. Rebuzzini,
Effects of a nonlinear perturbation on dynamical tunneling in cold atoms,
Phys. Rev. E \textbf{68}, 036221 (2003).

\bibitem{Got19}
H. Gothe, T. Valenzuela, M. Cristiani, and J. Eschner,
Optical bistability and nonlinear dynamics by saturation of cold Yb atoms in a cavity,
Phys. Rev. A \textbf{99}, 013849 (2019).

\bibitem{Par12}
V. Parigi, E. Bimbard, J. Stanojevic, A. J. Hilliard,
F. Nogrette, R. Tualle-Brouri, A. Ourjoumtsev, and P. Grangier,
Observation and Measurement of Interaction-Induced Dispersive Optical Nonlinearities in an Ensemble of Cold Rydberg Atoms,
Phys. Rev. Lett. \textbf{109}, 233602 (2012)

\bibitem{Ald05}
J. S. Aldridge and A. N. Cleland,
Noise-enabled precision measurement of a Duffing nanomechanical resonator,
Phys. Rev. Lett. \textbf{94}, 156403 (2005).


\bibitem{Sid04}
I. Siddiqi, R. Vijay, F. Pierre, C. M. Wilson, M. Metcalfe, C. Rigetti, L. Frunzio, and M. H. Devoret,
RF-driven Josephson bifurcation amplifier for quantum measurement,
Phys. Rev. Lett. \textbf{93}, 207002 (2004).

\bibitem{Sid06}
I. Siddiqi, R. Vijay, M. Metcalfe, E. Boaknin, L. Frunzio, R. J. Schoelkopf, and M. H. Devoret,
Dispersive measurements of superconducting qubit coherence with a fast latching readout,
Phys. Rev. B \textbf{73}, 054510 (2006).

\bibitem{Man07}
V. E. Manucharyan, E. Boaknin, M. Metcalfe, R. Vijay, I. Siddiqi, and M. Devoret,
Microwave bifurcation of a Josephson junction: Embedding-circuit requirements,
Phys. Rev. B \textbf{76}, 014524 (2007).

\bibitem{Sid09}
R. Vijay, M. H. Devoret, I. Siddiqi,
Invited Review Article: The Josephson bifurcation amplifier,
Rev. Sci. Instrum. \textbf{80}, 111101 (2009).


\bibitem{Chen23New}
Q.-M. Chen, M. Fischer, Y. Nojiri, {\it et al},
Quantum behavior of the Duffing oscillator at the dissipative phase transition,
Nat. Commun. \textbf{14}, 2896 (2023).

\bibitem{Mar21New}
A. D. Maris, B. Pokharel, S. G. Seshachallam, M. Z. R. Misplon, and A. K. Pattanayak,
Chaos in the quantum Duffing oscillator in the semiclassical regime under parametrized dissipation,
Phys. Rev. E \textbf{104}, 024206 (2021).



\bibitem{Dyk05}
M. I. Dykman, I. B. Schwartz, and M. Shapiro,
Scaling in activated escape of underdamped systems,
Phys. Rev. E \textbf{72}, 021102 (2005).


	\bibitem{Dyk11n}
	M. I. Dykman, M. Marthaler, and V. Peano,
	Quantum heating of a parametrically modulated oscillator: Spectral signatures,
	Phys. Rev. A \textbf{83}, 052115 (2011).
	
\bibitem{Dyk07}
M. I. Dykman,
Critical exponents in metastable decay via quantum activation,
Phys. Rev. E \textbf{75}, 011101 (2007).

\bibitem{Guo10}
L. Z. Guo, Z. G. Zheng, and X. -Q. Li,
Quantum dynamics of mesoscopic driven Duffing oscillators,
EPL \textbf{90}, 10011 (2010).

\bibitem{Guo11}
L. Z. Guo, Z. G. Zheng, X. -Q. Li, and Y. J. Yan,
Dynamic quantum tunneling in mesoscopic driven Duffing oscillators,
Phys. Rev. E \textbf{84}, 011144 (2011).

\bibitem{Ros24}
C. A. Rosiek, M. Rossi, A. Schliesser, and A. S. S{\o} rensen,
Quadrature Squeezing Enhances Wigner Negativity in a Mechanical Duffing Oscillator,
PRX Quantum \textbf{5}, 030312 (2024).

\bibitem{Coc24}
M. Coccolo, M. A.F. Sanju\'{a}n, Transmitted resonance in a coupled system,
Commun. Nonlinear Sci. \textbf{135}, 108068 (2024).



\bibitem{Pea10n}
V. Peano, and M. Thorwart,
Quasienergy description of the driven Jaynes-Cummings model,
Phys. Rev. B \textbf{82}, 155129 (2010).

\bibitem{Lem15n}
M.-A. Lemonde, A. A. Clerk,
Real photons from vacuum fluctuations in optomechanics: The role of polariton interactions,
Phys. Rev. A \textbf{91}, 033836 (2015).


\bibitem{Ser07}
I. Serban, and F. K. Wilhelm,
Dynamical Tunneling in Macroscopic Systems,
Phys. Rev. Lett. \textbf{99}, 137001 (2007).

\bibitem{Dyk06}
M. Marthaler, and M. I. Dykman,
Switching via quantum activation: A parametrically modulated oscillator,
Phys. Rev. A \textbf{73}, 042108 (2006).



\end{thebibliography}

\end{document}